\documentclass[twocolumn,english,aps,prd,nofootinbib]{revtex4-2}
\usepackage{lmodern}
\usepackage[T1]{fontenc}
\usepackage[latin9]{inputenc}
\usepackage{xcolor}
\usepackage{babel}
\usepackage{amsmath}
\usepackage{amssymb}
\usepackage{graphicx}
\usepackage[unicode=true]
 {hyperref}

\makeatletter

\usepackage{enumerate}

\makeatother

\begin{document}
\title{Non-triviality of an asymptotically flat vacuum spacetime in pure
$R^{2}$ gravity}
\author{Hoang Ky Nguyen$\,$}
\email[\ \ ]{hoang.nguyen@ubbcluj.ro}

\affiliation{Department of Physics, Babe\c{s}--Bolyai University, Cluj-Napoca
400084, Romania}
\date{May 28, 2024}
\begin{abstract}
\noindent \vskip2ptIn Phys.$\,$Rev.$\,$D \textbf{107}, 104008 (2023)
we reported a novel exact closed-form solution which describes asymptotically
flat spacetimes in pure $\mathcal{R}^{2}$ gravity. The solution is
Ricci scalar flat, viz. $\mathcal{R}\equiv0$ everywhere. Whereas
any metric with a null Ricci scalar would \emph{trivially} satisfy
the $\mathcal{R}^{2}$ vacuo field equation, $\mathcal{R}\left(\mathcal{R}_{\mu\nu}-\frac{1}{4}g_{\mu\nu}\,\mathcal{R}\right)+g_{\mu\nu}\,\square\,\mathcal{R}-\nabla_{\mu}\nabla_{\nu}\mathcal{R}=0$,
in this article, we shall show that our solution satisfies a ``stronger''
version of the $\mathcal{R}^{2}$ vacuo field equation, viz. $\mathcal{R}_{\mu\nu}-\frac{1}{4}g_{\mu\nu}\,\mathcal{R}+\mathcal{R}^{-1}\left(g_{\mu\nu}\,\square\,\mathcal{R}-\nabla_{\mu}\nabla_{\nu}\mathcal{R}\right)=0$,
despite the term $\mathcal{R}^{-1}$ being \emph{singular}. Even though
$\mathcal{R}$ identically vanishes, for our solution, the combinations
$\,\mathcal{R}^{-1}\,\nabla_{\mu}\nabla_{\nu}\mathcal{R}\,$ and $\,\mathcal{R}^{-1}\,\square\,\mathcal{R}\,$
are \emph{free of singularity}. This exceptional property sets our
solution apart from the set of null-Ricci-scalar metrics and makes
it a genuinely\emph{ non-trivial} solution. We further demonstrate
that, as a member of a larger class of asymptotically de Sitter metrics,
our solution is resilient against perturbations in the scalar curvature
at largest distances, making it relevant for physical situations where
the background deviates from asymptotic flatness.
\end{abstract}
\maketitle

\section{\label{sec:Motivation}Motivation}

In the study of gravitation, several exact solutions have been identified
for General Relativity (GR) \citep{solution-GR}. However, exact solutions
in modified theories of gravity are much rarer, as these theories
are more technically involved. Among the modifications to GR, the
family of $f(\mathcal{R})$ actions, with $\mathcal{R}$ being the
Ricci scalar curvature, put forth by Buchdahl \citep{Buchdahl-1970}
has been a popular arena for exploration, with important consequences
in early-time and late-time cosmology \citep{reviews}. A number of
exact solutions have been identified for $f(\mathcal{R})$ gravity
\citep{solutions-fR,Shankaranarayanan-2018}.\vskip4pt

The pure $\mathcal{R}^{2}$ theory is a simple extension beyond the
Einstein-Hilbert action and was first considered by Buchdahl in the
early 1960s as a prototype of higher-order gravity \citep{Buchdahl-1962}.
Recently, pure $\mathcal{R}^{2}$ gravity has regained interest as
both a classical field theory and a renormalizable quantum theory
of gravity \citep{AlvarezGaume-2015}. This theory is unique in being
both ghost-free and scale invariant \citep{Luest-2015-fluxes}. Its
action $(2\kappa)^{-1}\int d^{4}x\sqrt{-g}\,\mathcal{R}^{2}$, where
$\kappa$ is a dimensionless parameter, actually affords a \emph{restricted}
scale invariance, which lies between a global scale invariance and
a conformal invariance \citep{Edery-2014}.\vskip4pt

The pure $\mathcal{R}^{2}$ field equation in vacuo is \vskip-8pt

\begin{equation}
\mathcal{R}\left(\mathcal{R}_{\mu\nu}-\frac{1}{4}g_{\mu\nu}\mathcal{R}\right)+g_{\mu\nu}\square\,\mathcal{R}-\nabla_{\mu}\nabla_{\nu}\mathcal{R}=0\label{eq:R2-field-eqn}
\end{equation}
which involves fourth-derivative terms of the $g_{\mu\nu}$ components,
via $\nabla_{\mu}\nabla_{\mu}\mathcal{R}$. Despite its complexity,
we have identified an exact analytical solution \citep{Nguyen-2022-Lambda0},
which built upon a previous work by Buchdahl \citep{Buchdahl-1962}
as well as another recent work of ours \citep{Nguyen-2022-Buchdahl}.
This solution, which we dubbed the \emph{special} Buchdahl-inspired
metric in \citep{Nguyen-2022-Lambda0}, describes asymptotically flat
spacetimes in pure $\mathcal{R}^{2}$ gravity. In this present paper,
we shall also refer it as the asymptotically flat Buchdahl-inspired
metric, interchangeably.\vskip4pt

Thanks to its closed analytical form, the solution permitted us to
conduct tests against a wide range of observations, as recently reported
in \citep{2024-tests}. Furthermore, we constructed a Kruskal-Szekeres
diagram and explores its physical implications, including the existence
of naked singularities and wormholes \citep{Nguyen-2022-Lambda0}.
Notably, we identified a novel type of Morris-Thorne-Buchdahl wormholes
\citep{2023-Traversable-WH} and, consequently, uncovered a new class
of Closed Timelike Curves in these wormholes \citep{2023-CTC,2024-timereverse}.
We also studied its axisymmetric extension and applied it to the M87{*}
shadow \citep{2023-axisym}.\vskip4pt

Our current paper aims to delve deeper and establish the \emph{non-triviality}
of the solution. There are a number of representations for the \emph{special}
Buchdahl-inspired metric, with the two most illuminating ones being
given below (with $d\Omega^{2}:=d\theta^{2}+\sin^{2}\theta\,d\varphi^{2}$).\vskip12pt

\emph{Representation \#1}, first derived in \citep{Nguyen-2022-Lambda0}:
\footnote{Note that we used a different set of notation for variables in \citep{Nguyen-2022-Lambda0}.}\small
\begin{align}
ds^{2} & =\left|1-\frac{r_{\text{s}}}{r}\right|^{\tilde{k}}\times\nonumber \\
 & \ \ \biggl\{-\left(1-\frac{r_{\text{s}}}{r}\right)dt^{2}+\frac{dr^{2}}{1-\frac{r_{\text{s}}}{r}}\frac{\rho^{4}(r)}{r^{4}}+\rho^{2}(r)d\Omega^{2}\biggr\}\label{eq:special-B-1}\\
\rho^{2}(r) & :=\frac{\zeta^{2}r_{\text{s}}^{2}\,\left|1-\frac{r_{\text{s}}}{r}\right|^{\zeta-1}}{\left(1-\text{sgn}\left(1-\frac{r_{\text{s}}}{r}\right)\left|1-\frac{r_{\text{s}}}{r}\right|^{\zeta}\right)^{2}}\label{eq:special-B-2}
\end{align}
\normalsize\vskip8pt

\emph{Representation \#2}, first reported in \citep{2023-Traversable-WH}:\small
\begin{align}
ds^{2} & =-\left|1-\frac{\zeta r_{\text{s}}}{r'}\right|^{\frac{\tilde{k}+1}{\zeta}-1}\left(1-\frac{\zeta r_{\text{s}}}{r'}\right)dt^{2}\nonumber \\
 & \ \ \ +\left|1-\frac{\zeta r_{\text{s}}}{r'}\right|^{\frac{\tilde{k}-1}{\zeta}+1}\left[\frac{dr'^{2}}{1-\frac{r_{\text{s}}}{r'}}+r'^{2}d\Omega^{2}\right]\label{eq:special-B-3}
\end{align}
\normalsize The two representations are related via a coordinate
transformation:
\begin{equation}
1-\frac{\zeta r_{\text{s}}}{r'}=\text{sgn}\left(1-\frac{r_{\text{s}}}{r}\right)\left|1-\frac{r_{\text{s}}}{r}\right|^{\zeta}
\end{equation}
In either representation, the metric contains two parameters, $r_{\text{s}}$
playing the role of a Schwarzschild radius, and $\tilde{k}$ a new
(Buchdahl) dimensionless parameter, and $\zeta:=\sqrt{1+3\tilde{k}^{2}}$.
The solution holds for all value of $r\in\mathbb{R}$ except at $r=0$
and $r=r_{\text{s}}$, where its Kretschmann invariant diverges, indicating
physical singularities \citep{Nguyen-2022-Lambda0}. \vskip4pt

One important question remains to be addressed in this paper. It was
found in Ref.$\ $\citep{Nguyen-2022-Lambda0} that the \emph{special}
Buchdahl-inspired metric, given in Eqs. \eqref{eq:special-B-1}--\eqref{eq:special-B-2}
or Eq. \eqref{eq:special-B-3}, has a vanishing Ricci scalar everywhere,
viz. $\mathcal{R}\equiv0\ \ \forall r$. It is self-evident that any
metric that is Ricci-scalar flat would automatically satisfy the field
equation \eqref{eq:R2-field-eqn}. Such a metric is, therefore, a
\emph{trivial} solution. There exists a broad range of such metrics,
subject to \emph{one} constraint $\mathcal{R}=0$, which is generally
insufficient to determine the metric $g_{\mu\nu}$, resulting in the
metric being under-determined.\vskip4pt

The purpose of our paper is to demonstrate that the \emph{special}
Buchdahl-inspired metric satisfies a ``stronger'' version of the $\mathcal{R}^{2}$
vacuo field equation, that is
\begin{equation}
\mathcal{R}_{\mu\nu}-\frac{1}{4}g_{\mu\nu}\mathcal{R}-\mathcal{R}^{-1}\Bigl(\nabla_{\mu}\nabla_{\nu}\mathcal{R}-g_{\mu\nu}\square\,\mathcal{R}\Bigr)=0\label{eq:R2-field-eqn-div}
\end{equation}
\emph{despite that $\mathcal{R}^{-1}$ is singular} (due to $\mathcal{R}\equiv0$).
This metric circumvents the singularity in $\mathcal{R}^{-1}$ by
having the term $\nabla_{\mu}\nabla_{\nu}\mathcal{R}$ vanish in such
a way that the ratio $\mathcal{R}^{-1}\nabla_{\mu}\nabla_{\nu}\mathcal{R}$
stays finite and equals $\mathcal{R}_{\mu\nu}$. The equality then
results in $\mathcal{R}^{-1}\square\,\mathcal{R}=\mathcal{R}=0$.
We present the proof of this in Section \ref{sec:Proof-of-non-triviality}.\vskip4pt

It is worth noting that the field equation in its ``stronger'' form
of \eqref{eq:R2-field-eqn-div} still retains sufficient constraints,
(that is, \emph{ten} coupled partial differential equations), along
with appropriate boundary conditions, to determine the \emph{ten}
symmetric components of the metric $g_{\mu\nu}$. The \emph{special}
Buchdahl-inspired metric is thus fully determined, constituting a
\emph{non-trivial} solution.\vskip4pt

We conclude in Section \ref{sec:Discussion}, based on our proof,
that the pure $\mathcal{R}^{2}$ action selects the \emph{special}
Buchdahl-inspired metric to be its \emph{asymptotically flat} vacuum,
instead of settling for an arbitrary null-Ricci-scalar metric despite
the latter automatically fulfilling to the field equation.

\section{\label{sec:Proof-of-non-triviality}Proof of non-triviality}

Directly computing the terms $\mathcal{R}^{-1}\nabla_{\mu}\nabla_{\nu}\mathcal{R}$
and $\mathcal{R}^{-1}\square\,\mathcal{R}$ from the metric \eqref{eq:special-B-1}--\eqref{eq:special-B-2}
(or \eqref{eq:special-B-3}) is problematic because $\mathcal{R}$,
$\nabla_{\mu}\nabla_{\nu}\mathcal{R}$, and $\square\,\mathcal{R}$
all vanish. However, we can overcome this difficulty by utilizing
another recent result. Specifically, we can exploit that fact that
the asymptotically flat Buchdahl-inspired metric is a special member
of a more general class of metrics, also inspired by Buchdahl \citep{Buchdahl-1962,Nguyen-2022-Buchdahl}.\vskip4pt

In Ref.$\ $\citep{Nguyen-2022-Buchdahl}, a \emph{general} class
of (Buchdahl-inspired) metrics that are asymptotically \emph{constant}
was derived. Each member of the class is a solution to the pure $\mathcal{R}^{2}$
vacuo field equation \eqref{eq:R2-field-eqn} and is expressible in
the following compact form

\small
\begin{align}
ds^{2} & =e^{k\int\frac{dr}{r\,q(r)}}\left\{ p(r)\Bigl[-\frac{q(r)}{r}dt^{2}+\frac{r}{q(r)}dr^{2}\Bigr]+r^{2}d\Omega^{2}\right\} \label{eq:B-metric}
\end{align}
\normalsize where the pair of functions $\{p(r),q(r)\}$ obey the
``evolution'' rules
\begin{align}
{\displaystyle \frac{dp}{dr}} & ={\displaystyle \frac{3k^{2}}{4\,r}\frac{p}{q^{2}}\ \ \ \text{and}\ \ \ \frac{dq}{dr}=\left(1-\Lambda\,r^{2}\right)p}\label{eq:evol}
\end{align}
and the Ricci scalar is given by 
\begin{equation}
\mathcal{R}=4\Lambda\,e^{-k\int\frac{dr}{r\,q(r)}}\label{eq:Ricci}
\end{equation}
Its Ricci scalar is \emph{non-constant} and approaches $4\Lambda$
at spatial infinity. \footnote{Note that the general Buchdahl-inspired metrics, with \emph{non-constant}
scalar curvature, defeat the generalized Lichnerowicz theorem as we
explained in \citep{Nguyen-2022-extension,Lich-theorem}.}\vskip4pt

When $\Lambda=0$ the evolution rules \eqref{eq:evol} become \emph{fully
soluble}. In Ref.$\ $\citep{Nguyen-2022-Lambda0} we solved these
rules for $\Lambda=0$ and obtained the metric given in \eqref{eq:special-B-1}--\eqref{eq:special-B-2}.
That is to say, the \emph{special} Buchdahl-inspired metric \eqref{eq:special-B-1}--\eqref{eq:special-B-2}
is \emph{the} $\Lambda=0$ member of the \emph{general} Buchdahl-inspired
metric \eqref{eq:B-metric}--\eqref{eq:Ricci}. Therefore, we can
first compute the terms $\mathcal{R}^{-1}\nabla_{\mu}\nabla_{\nu}\mathcal{R}$
and $\mathcal{R}^{-1}\square\,\mathcal{R}$ for the \emph{general}
Buchdahl-inspired metric, which is free of singularity, then send
$\Lambda$ to zero to obtain the terms for the \emph{special} Buchdahl-inspired
metric.\vskip4pt

For a static spherically symmetric metric, only the four diagonal
components of the field equation are of our interest. However, the
$\phi\phi$ component is trivially proportional the $\theta\theta$
component by the factor $\sin^{2}\theta$; i.e., $\mathcal{R}_{\phi\phi}=\mathcal{R}_{\theta\theta}\sin^{2}\theta$
and $\Gamma_{\phi\phi}^{r}=\Gamma_{\theta\theta}^{r}\sin^{2}\theta$.
Thus we only need to check the three components $tt$-, $rr$-, and
$\theta\theta$- of the $\mathcal{R}^{2}$ field equation. The task
is non-trivial because of the cross-dependence between $p(r)$ and
$q(r)$ per Eq. \eqref{eq:evol}. \vskip4pt

Since $\mathcal{R}$ is a function of $r$ only, we have \footnote{Recall that for a scalar field $\phi$: $\nabla_{\mu}\nabla_{\nu}\phi=\partial_{\mu}\partial_{\nu}\phi-\Gamma_{\mu\nu}^{\lambda}\partial_{\lambda}\phi$.}
\begin{equation}
\nabla_{\mu}\nabla_{\nu}\mathcal{R}=\partial_{\mu}\partial_{\nu}\mathcal{R}-\Gamma_{\mu\nu}^{r}\,\partial_{r}\mathcal{R}
\end{equation}
The diagonal components of this tensor read
\begin{align}
\nabla_{t}\nabla_{t}\mathcal{R} & =-\Gamma_{tt}^{r}\mathcal{R}'\label{eq:00-eqn}\\
\nabla_{r}\nabla_{r}\mathcal{R} & =-\Gamma_{rr}^{r}\mathcal{R}'+\mathcal{R}''\label{eq:11-eqn}\\
\nabla_{\theta}\nabla_{\theta}\mathcal{R} & =-\Gamma_{\theta\theta}^{r}\mathcal{R}'\label{eq:22-eqn}\\
\nabla_{\phi}\nabla_{\phi}\mathcal{R} & =\nabla_{\theta}\nabla_{\theta}\mathcal{R}\ \sin^{2}\theta\label{eq:33-eqn}
\end{align}

The \emph{general} Buchdahl-inspired metric \eqref{eq:special-B-1}
involves implicit functions of $p(r)$ and $q(r)$: \footnote{Note that $p(r)$ and $q(r)$ are given implicitly via \eqref{eq:evol},
and depend on $\Lambda$.}
\begin{equation}
\begin{cases}
\ g_{tt} & =-e^{k\int\frac{dr}{r\,q(r)}}\,\frac{p\,q}{r}\\
\ g_{rr} & =\ \,\ e^{k\int\frac{dr}{r\,q(r)}}\,\frac{p\,r}{q}\\
\ g_{\theta\theta} & =\ \ \,e^{k\int\frac{dr}{r\,q(r)}}\,r^{2}\\
\ g_{\varphi\varphi} & =\ \ \,e^{k\int\frac{dr}{r\,q(r)}}\,r^{2}\sin^{2}\theta
\end{cases}
\end{equation}
Its Ricci tensor components hence involve second-derivative terms
$p''(r)$ and $q''(r)$. Likewise, the covariant derivatives of its
Ricci scalar are of fourth-order, $p''''(r)$ and $q''''(r)$. By
applying the evolution rules \eqref{eq:evol} recursively, we can
reduce the higher-derivative terms to algebraic functions of $p$
and $q$. The details are presented in the Appendix of this paper.\vskip8pt

The non-vanishing components of the Ricci tensor are
\begin{equation}
\begin{cases}
\ \mathcal{R}_{tt} & =-k\,\frac{q^{2}-\left(k+rp\right)q-\frac{3}{4}k^{2}}{2r^{4}q}-\frac{\left(2q+k\right)p}{2r}\,\Lambda\\
\ \mathcal{R}_{rr} & =k\,\frac{3q^{2}+\left(3k+rp\right)q+\frac{3}{4}k^{2}}{2r^{2}q^{3}}+\frac{\left(2q-k\right)rp}{2q^{2}}\,\Lambda\\
\ \mathcal{R}_{\theta\theta} & =-k\,\frac{2q+k}{2rpq}+r^{2}\Lambda\\
\ \mathcal{R}_{\varphi\varphi} & =\left[-k\,\frac{2q+k}{2rpq}+r^{2}\Lambda\right]\,\sin^{2}\theta
\end{cases}
\end{equation}
from which, the Ricci scalar is
\begin{equation}
\mathcal{R}:=g^{\mu\nu}\mathcal{R}_{\mu\nu}=4\Lambda\,e^{-k\int\frac{dr}{r\,q}}\label{eq:valid-5}
\end{equation}
\vskip8pt

The non-vanishing components of $\mathcal{\mathcal{R}}^{-1}\nabla_{\mu}\nabla_{\nu}\mathcal{R}$
are
\begin{equation}
\begin{cases}
\ \mathcal{R}^{-1}\nabla_{t}\nabla_{t}\mathcal{R} & =-k\,\frac{q^{2}-\left(k+rp\right)q-\frac{3}{4}k^{2}}{2r^{4}q}-\frac{kp}{2r}\,\Lambda\\
\ \mathcal{R}^{-1}\nabla_{r}\nabla_{r}\mathcal{R} & =k\,\frac{3q^{2}+\left(rp+3k\right)q+\frac{3}{4}k^{2}}{2r^{2}q^{3}}-\frac{k\,rp}{2q^{2}}\,\Lambda\\
\ \mathcal{R}^{-1}\nabla_{\theta}\nabla_{\theta}\mathcal{R} & =-k\,\frac{2q+k}{2rpq}\\
\ \mathcal{R}^{-1}\nabla_{\varphi}\nabla_{\varphi}\mathcal{R} & =-k\,\frac{2q+k}{2rpq}\,\sin^{2}\theta
\end{cases}
\end{equation}
from which, it is obvious that
\begin{equation}
\mathcal{R}^{-1}\,\square\,\mathcal{R}:=g^{\mu\nu}\left(\mathcal{R}^{-1}\nabla_{\mu}\nabla_{\nu}\mathcal{R}\right)\equiv0\ \ \ \forall r\label{eq:valid-10}
\end{equation}
\vskip8pt

Although $\mathcal{R_{\mu\nu}\neq\mathcal{R}}^{-1}\nabla_{\mu}\nabla_{\nu}\mathcal{R}$
in general, the following equalities hold: \footnote{Note that the functions $p$ and $q$ in these latest relations obey
the evolution rules with $\Lambda$ set equal zero, namely $\frac{dp}{dr}=\frac{3k^{2}}{4\,r}\frac{p}{q^{2}}$
and $\frac{dq}{dr}=p$ which are fully soluble as provided in Ref.$\ $\citep{Nguyen-2022-Lambda0},
but their concrete expressions are not relevant for our discussion
here. }
\begin{equation}
\begin{cases}
\ {\displaystyle \lim_{\Lambda\rightarrow0}\mathcal{R}^{-1}\nabla_{t}\nabla_{t}\mathcal{R}\,=\lim_{\Lambda\rightarrow0}\mathcal{R}_{tt}} & =-k\,\frac{q^{2}-\left(k+rp\right)q-\frac{3}{4}k^{2}}{2r^{4}q}\\
\ {\displaystyle \lim_{\Lambda\rightarrow0}\mathcal{R}^{-1}\nabla_{r}\nabla_{r}\mathcal{R}=\lim_{\Lambda\rightarrow0}\mathcal{R}_{rr}} & =k\,\frac{3q^{2}+\left(rp+3k\right)q+\frac{3}{4}k^{2}}{2r^{2}q^{3}}\\
\ {\displaystyle \lim_{\Lambda\rightarrow0}\mathcal{R}^{-1}\nabla_{\theta}\nabla_{\theta}\mathcal{R}=\lim_{\Lambda\rightarrow0}\mathcal{R}_{\theta\theta}} & =-k\,\frac{2q+k}{2rpq}\\
\ {\displaystyle \lim_{\Lambda\rightarrow0}\mathcal{R}^{-1}\nabla_{\phi}\nabla_{\phi}\mathcal{R}=\lim_{\Lambda\rightarrow0}\mathcal{R}_{\phi\phi}} & =-k\,\frac{2q+k}{2rpq}\sin^{2}\theta
\end{cases}\label{eq:valid-11}
\end{equation}
Thus
\begin{equation}
\lim_{\Lambda\rightarrow0}\mathcal{R}^{-1}\nabla_{\mu}\nabla_{\nu}\mathcal{R}=\lim_{\Lambda\rightarrow0}\mathcal{R}_{\mu\nu}.\label{eq:valid-12}
\end{equation}
In addition, by virtue of \eqref{eq:valid-5}
\begin{equation}
{\displaystyle \lim_{\Lambda\rightarrow0}\mathcal{R}}=0.\label{eq:valid-13}
\end{equation}

The relation \eqref{eq:valid-12} is remarkable in that, despite $\mathcal{R}^{-1}$
being singular in the $\Lambda\rightarrow0$ limit, the left hand
side of \eqref{eq:valid-12} is free of singularity. Together with
\eqref{eq:valid-10} and \eqref{eq:valid-13}, this relation establishes
that the \emph{special} Buchdahl-inspired metric, namely, the general
Buchdahl-inspired metric with $\Lambda=0$, satisfies both equations
below
\begin{equation}
\mathcal{R}=0,
\end{equation}
and 
\begin{equation}
\mathcal{R}_{\mu\nu}-\frac{1}{4}g_{\mu\nu}\mathcal{R}=\mathcal{R}^{-1}\left(\nabla_{\mu}\nabla_{\nu}\mathcal{R}-g_{\mu\nu}\square\,\mathcal{R}\right).
\end{equation}
QED.

\section{\label{sec:Discussion}Discussion: $\ $Viability against $\mathcal{O}\left(\Lambda\right)$
perturbations}

\begin{figure}[t]
\noindent \begin{centering}
$\ \ \ $\includegraphics[scale=0.75]{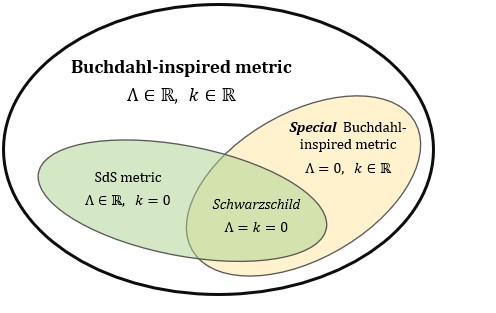}\vskip10pt
\par\end{centering}
\noindent \begin{centering}
$\ \ \ $\includegraphics[scale=0.75]{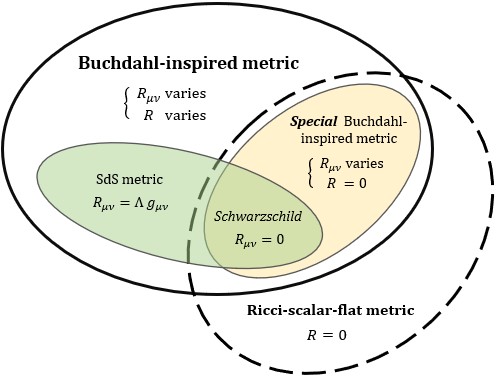}
\par\end{centering}
\caption{\label{fig:relation}Relation of the Buchdahl-inspired metric and
others. The \emph{special} Buchdahl-inspired metric is asymptotically
flat, occupying the intersection of the Buchdahl-inspired metric family
and the null-Ricci-scalar metric family.}
\end{figure}

Stated in a slightly different perspective, the vacuo field equation
\eqref{eq:R2-field-eqn} can be ``factorized'' into
\begin{equation}
\mathcal{R}\left[\mathcal{R}_{\mu\nu}-\frac{1}{4}g_{\mu\nu}\mathcal{R}+\mathcal{R}^{-1}\left(g_{\mu\nu}\square\,\mathcal{R}-\nabla_{\mu}\nabla_{\nu}\mathcal{R}\right)\right]=0\label{eq:1b}
\end{equation}
Therefore it corresponds to two branches of solutions:
\begin{enumerate}
\item A branch of \emph{trivial} solutions, namely, the set of null-Ricci-scalar
metrics
\begin{equation}
\mathcal{R}=0\label{eq:1c}
\end{equation}
\item A branch of \emph{non-trivial} solutions, satisfying the ``stronger''
version 
\begin{equation}
\mathcal{R}_{\mu\nu}-\frac{1}{4}g_{\mu\nu}\mathcal{R}+\mathcal{R}^{-1}\left(g_{\mu\nu}\square\,\mathcal{R}-\nabla_{\mu}\nabla_{\nu}\mathcal{R}\right)=0\tag{\ref{eq:R2-field-eqn-div}}
\end{equation}
which comprises of the \emph{general} class of Buchdahl-inspired solutions
given in \eqref{eq:B-metric}--\eqref{eq:Ricci} with $\Lambda\in\mathbb{R}$.\vskip4pt
\end{enumerate}
The branch of trivial solutions and the branch of non-trivial solutions
intersect with each other, with the \emph{special} Buchdahl-inspired
metric belonging simultaneously to both branches. This interesting
property can be visualized in the Venn diagrams shown in Figure \ref{fig:relation}.\vskip4pt

Finding metrics that are merely Ricci-scalar-flat is a relatively
simple task, as they are subject to only \emph{one} single constraint,
$\mathcal{R}=0$, while a 4-dimensional metric has at least two degrees
of freedom. The most generic form for a static spherically symmetric
metric in the Schwarzschild coordinate system, viz.
\begin{equation}
ds^{2}=-A(r)dt^{2}+B(r)dr^{2}+r^{2}d\Omega^{2}\label{eq:viab-1}
\end{equation}
contains two unknown functions $\{A(r),B(r)\}$ against one equation,
$\mathcal{R}=0$, and thus is under-determined. Even when the additional
requirement of asymptotic flatness is imposed, it only somewhat restricts
$\{A(r),\,B(r)\}$ but these two functions remain under-determined.
A couple of null-Ricci-scalar metrics that are asymptotically flat
have been given in \citep{Shankaranarayanan-2018}:
\begin{equation}
ds^{2}=-dt^{2}+\frac{dr^{2}}{1-\frac{r_{s}}{r}}+r^{2}d\Omega^{2}\label{eq:viab-2}
\end{equation}
and
\begin{equation}
ds^{2}=-\left(1-\frac{r_{\text{s}}}{r}+\frac{Q^{2}}{r^{2}}\right)dt^{2}+\frac{dr^{2}}{1-\frac{r_{s}}{r}+\frac{Q^{2}}{r^{2}}}+r^{2}d\Omega^{2}\label{eq:viab-3}
\end{equation}
They both have $\mathcal{R}_{\mu\nu}\neq0$ yet $\mathcal{R}=0$.
These metrics belong to the branch of trivial vacuum solutions for
the pure $\mathcal{R}^{2}$ action. More generally, any null-Ricci-scalar
metric would be a trivial vacuum solution to the $f(\mathcal{R})$
theories that are ``non-pathological''. For example, the $\mathcal{R}^{n}$
theory with $n>1$ has the vacuo field equation viz. $n\mathcal{R}^{n-1}\mathcal{R}_{\mu\nu}-\frac{1}{2}\mathcal{R}^{n}g_{\mu\nu}+n\left(g_{\mu\nu}\square-\nabla_{\mu}\nabla_{\nu}\right)\mathcal{R}^{n-1}=0$
which automatically admits any null-Ricci-scalar metric as a solution.\vskip4pt

Yet, Figure \ref{fig:relation} highlights a distinct property of
the asymptotically flat Buchdahl-inspired metric within the set of
the null-Ricci-scalar metrics. By occupying the intersection of the
two solution branches, this metric can smoothly transition into a
member of the nontrivial branch of the general Buchdahl-inspired metric
as the parameter $\Lambda$ departs from zero. This remarkable flexibility
enables the asymptotically flat Buchdahl-inspired metric to \emph{absorb
$\mathcal{O}(\Lambda)$ perturbations} when the background experiences
slight scalar curvature variations at large distances. On the other
hand, any other arbitrary solution with $\mathcal{R}=0$ can only
move within the trivial branch and cannot evolve into a nontrivial
solution with $\Lambda\neq0$. \vskip4pt

Another way to put it is that when a non-vanishing scalar curvature
$4\Lambda$ is present at large distances, the pure $\mathcal{R}^{2}$
action allows for a general Buchdahl-inspired metric as its vacuum
solution. However, \emph{deep in the bulk} where $\left|\Lambda r^{2}\right|\ll1$,
the effect of $\Lambda$ becomes negligible in the evolution rules
\eqref{eq:evol}, and the asymptotically flat Buchdahl-inspired metric
remains an accurate description of the spacetime in that region. In
contrast, an arbitrary null-Ricci-scalar flat metric is inadequate
for describing such a situation.

\section{Conclusions}

In Section \ref{sec:Proof-of-non-triviality}, we demonstrated that
the combination $\mathcal{R}^{-1}\nabla_{\mu}\nabla_{\nu}\mathcal{R}$
of the \emph{general} Buchdahl-inspired metric, derived in \citep{Nguyen-2022-Buchdahl},
is well-behaved for all values of $\Lambda$, a parameter that characterizes
the scalar curvature of the metric at largest distances. In the limit
of $\Lambda\rightarrow0$, the Ricci scalar identically vanishes everywhere.
However, the following equalities hold
\[
\lim_{\Lambda\rightarrow0}\mathcal{R}^{-1}\,\square\,\mathcal{R}=0\ \ \ \forall r\tag{\ref{eq:valid-10}}
\]
and\vskip-18pt
\[
\lim_{\Lambda\rightarrow0}\mathcal{R}^{-1}\nabla_{\mu}\nabla_{\nu}\mathcal{R}=\lim_{\Lambda\rightarrow0}\mathcal{R}_{\mu\nu}\tag{\ref{eq:valid-12}}
\]
As a result, the asymptotically flat Buchdahl-inspired metric, which
is a general Buchdahl-inspired metric when $\Lambda=0$, has well-defined
$\mathcal{R}^{-1}\nabla_{\mu}\nabla_{\nu}\mathcal{R}$ and $\mathcal{R}^{-1}\,\square\,\mathcal{R}$,
despite the term $\mathcal{R}^{-1}$ being singular. By virtue of
the equalities \eqref{eq:valid-10} and \eqref{eq:valid-12}, the
asymptotically flat Buchdahl-inspired metric satisfies the ``stronger''
version of the field equation
\[
\mathcal{R}_{\mu\nu}-\frac{1}{4}g_{\mu\nu}\mathcal{R}+\mathcal{R}^{-1}\left(g_{\mu\nu}\square\,\mathcal{R}-\nabla_{\mu}\nabla_{\nu}\mathcal{R}\right)=0\tag{\ref{eq:R2-field-eqn-div}}
\]
while avoiding a delicate issue with the degenerate $\mathcal{R}\equiv0$
everywhere.\vskip4pt

Unlike the $\mathcal{R}=0$ equation which leaves a null-Ricci-scalar
metric under-determined, the field equation in the ``stronger'' version
sufficiently determines the asymptotically flat Buchdahl-inspired
metric, making it a \emph{non-trivial} solution. The metric is also
resilient against $\mathcal{O}(\Lambda)$ perturbations, as discussed
in Section \ref{sec:Discussion}.\vskip8pt

Given the potential applications of the asymptotically flat Buchdahl-inspired
metric, as highlighted in previous studies \citep{2023-axisym,2023-Traversable-WH,2023-CTC,2024-tests,2024-timereverse},
and our recent investigation into its relevance to Sgr A{*} and its
star system \citep{Tao-2024}, we anticipate that this \emph{non-trivial}
metric, as established in our present paper, will continue to stimulate
further exploration and interest in $\mathcal{R}^{2}$ gravity.\vskip8pt

\textbf{{\emph{Outlook}}}---One
interesting extension is the Lagrangian $\sqrt{-g}\,\bigl[\mathcal{R}^{n}+\alpha\left(\square\mathcal{R}\right)^{m}\bigr]$
where $\alpha$ is a constant and $\square\equiv\nabla^{\mu}\nabla_{\mu}$.
This type of higher-order gravity belongs to the class $\sqrt{-g}\,F(\mathcal{R},\square\mathcal{R},...,\square^{m}\mathcal{R})$
introduced by Schmidt in \citep{Schmidt-1990-a}; it has interesting
cosmological consequences first explored in \citep{Schmidt-1990-b}.
Being Ricci scalar flat, the metric \eqref{eq:special-B-1} or \eqref{eq:special-B-3}
is automatically a (trivial) vacuum solution to the field equation
of the former Lagrangian provided that $m,\,n>1$. A worthwhile question
would be whether this metric also constitutes a \emph{non-trivial} solution, similar to the case considered herein, to that field equation
with $\alpha\neq0$ and a specific choice of $m=n=2$. As our current
proof was tailored for pure $\mathcal{R}^{2}$ gravity by relying
on an explicit expression of its asymptotically de Sitter vacuum solution
\citep{Nguyen-2022-Buchdahl}, this inquiry merits future exploration.
\begin{acknowledgments}
The author thanks the anonymous referee for insightful feedback which
opens up further explorations. Helpful commentaries from Tao Zhu,
Richard Shurtleff, Tiberiu Harko, Dieter L\"ust, Sergei Odintsov,
and Mustapha Azreg-A\"inou are acknowledged.
\end{acknowledgments}

\begin{center}
-----------------$\infty$-----------------
\par\end{center}

\appendix

\section{$\ $Detailed calculations of tensor components}

This Appendix shows the intermediate steps in the calculation of the
relevant tensor components in support of the Proof in Section \ref{sec:Proof-of-non-triviality}.
\vskip8pt

\textbf{Step 1.---}We start by considering the line element
\begin{equation}
ds^{2}=-e^{\nu(r)}dt^{2}+e^{\lambda(r)}dr^{2}+e^{\mu(r)}\left(d\theta^{2}+\sin^{2}\theta d\phi^{2}\right)
\end{equation}
in which, by virtue of \eqref{eq:B-metric} \small
\begin{align}
\nu(r) & :=\ln\left(f(r)\,\frac{p(r)\,q(r)}{r}\right)\label{eq:nu-def}\\
\lambda(r) & :=\ln\left(f(r)\,\frac{p(r)\,r}{q(r)}\right)\\
\mu(r) & :=\ln\left(f(r)\,r^{2}\right)
\end{align}
\normalsize which $f(r)$ unspecified for the moment. The relevant
Christoffel symbols and Ricci tensor components are\small
\begin{align}
\Gamma_{tt}^{r}\,e^{\lambda-\nu} & =\frac{\nu'}{2}\label{eq:Chris-00}\\
\Gamma_{\theta\theta}^{r}\,e^{\lambda-\mu} & =-\frac{\mu'}{2}\label{eq:Chris-22}\\
\Gamma_{rr}^{r} & =\frac{\lambda'}{2}\label{eq:Chris-11}
\end{align}
\begin{align}
\mathcal{R}_{tt}e^{\lambda-\nu} & =\frac{\nu''}{2}+\frac{\nu'^{2}}{4}-\frac{\nu'\lambda'}{4}+\frac{\nu'\mu'}{2}\\
-\mathcal{R}_{\theta\theta}e^{\lambda-\mu} & =-e^{\lambda-\mu}+\frac{\mu''}{2}+\frac{\mu'^{2}}{2}+\frac{\nu'\mu'}{4}-\frac{\lambda'\mu'}{4}\\
-\mathcal{R}_{rr} & =\frac{\nu''}{2}+\frac{\nu'^{2}}{4}+\mu''+\frac{\mu'^{2}}{2}-\frac{\nu'\lambda'}{4}-\frac{\lambda'\mu'}{2}\label{eq:R-rr}
\end{align}
\normalsize The Ricci scalar and its derivatives with respect to
$r$ expressed in terms of $\nu$, $\lambda$, $\mu$ are

\footnotesize
\begin{align}
\mathcal{R} & =-e^{-\lambda}\Bigl(\nu''+\frac{\nu'^{2}}{2}-\frac{\nu'\lambda'}{2}+\nu'\mu'-\lambda'\mu'+2\mu''+\frac{3}{2}\mu'^{2}\Bigr)\ \ \ \ \ \ \nonumber \\
 & \ \ \ \ +2e^{-\mu}\\
\mathcal{R}' & =e^{-\lambda}\Bigl(-\nu'''-\nu'\nu''-\mu'\nu''+\frac{3}{2}\lambda'\nu''+\frac{1}{2}\lambda'\nu'^{2}\nonumber \\
 & -\mu''\nu'+\lambda'\mu'\nu'+\frac{\lambda''\nu'}{2}-\frac{1}{2}\lambda'^{2}\nu'-2\mu'''-3\mu'\mu''\nonumber \\
 & +3\lambda'\mu''+\frac{3}{2}\lambda'\mu'^{2}+\lambda''\mu'-\lambda'^{2}\mu'\Bigr)-2e^{-\mu}\mu'
\end{align}
\begin{align}
\mathcal{R}'' & =e^{-\lambda}\Bigl(-\nu''''-\nu'\nu'''-\mu'\nu'''+\frac{5}{2}\lambda'\nu'''-\nu''^{2}\nonumber \\
 & +2\lambda'\nu'\nu''-2\mu''\nu''+2\lambda'\mu'\nu''+2\lambda''\nu''-2\lambda'^{2}\nu''\nonumber \\
 & +\frac{1}{2}\lambda''\nu'^{2}-\frac{1}{2}\lambda'^{2}\nu'^{2}-\mu'''\nu'+2\lambda'\mu''\nu'+\lambda''\mu'\nu'\nonumber \\
 & -\lambda'^{2}\mu'\nu'+\frac{1}{2}\lambda'''\nu'-\frac{3}{2}\lambda'\lambda''\nu'+\frac{1}{2}\lambda'^{3}\nu'-2\mu''''\nonumber \\
 & -3\mu'\mu'''+5\lambda'\mu'''-3\mu''^{2}+6\lambda'\mu'\mu''+4\lambda''\mu''-4\lambda'^{2}\mu''\nonumber \\
 & +\frac{3}{2}\lambda''\mu'^{2}-\frac{3}{2}\lambda'^{2}\mu'^{2}+\lambda'''\mu'-3\lambda'\lambda''\mu'+\lambda'^{3}\mu'\Bigr)\nonumber \\
 & -2e^{-\mu}\bigl(\mu''-\mu'^{2}\bigr)\label{eq:ddR}
\end{align}
\normalsize\vskip8pt

\textbf{Step 2.---}We further equate
\begin{equation}
f(r):=\exp\left(k\int\frac{dr}{r\,q(r)}\right)\label{eq:f-def}
\end{equation}
while leaving $p(r)$ and $q(r)$ \emph{unspecified} at the moment.
\vskip6pt Next, we find the derivatives of $\nu$, $\lambda$, $\mu$
with respect to $r$ in terms of $\{p(r),\ q(r)\}$ and their derivatives.
From \eqref{eq:nu-def}--\eqref{eq:f-def}, the results are (Notice
that $\lambda''''$ will not be needed):

\footnotesize
\begin{align}
\nu' & =\frac{q'}{q}+\frac{p'}{p}+\frac{k}{rq}-\frac{1}{r}\label{eq:b-1}\\
\lambda' & =-\frac{q'}{q}+\frac{p'}{p}+\frac{k}{rq}+\frac{1}{r}\\
\mu' & =\frac{2}{r}+\frac{k}{rq}
\end{align}
\begin{align}
\nu'' & =\frac{q''}{q}-\frac{q'^{2}}{q^{2}}-\frac{kq'}{rq^{2}}+\frac{p''}{p}-\frac{p'^{2}}{p^{2}}-\frac{k}{r^{2}q}+\frac{1}{r^{2}}\\
\lambda'' & =-\frac{q''}{q}+\frac{q'^{2}}{q^{2}}-\frac{kq'}{rq^{2}}+\frac{p''}{p}-\frac{p'^{2}}{p^{2}}-\frac{k}{r^{2}q}-\frac{1}{r^{2}}\\
\mu'' & =-\frac{kq'}{rq^{2}}-\frac{k}{r^{2}q}-\frac{2}{r^{2}}
\end{align}
\begin{align}
\nu''' & =\frac{q'''}{q}-\frac{3q'q''}{q^{2}}-\frac{kq''}{rq^{2}}+\frac{2q'^{3}}{q^{2}}+\frac{2kq'^{2}}{rq^{3}}+\frac{2kq'}{r^{2}q^{2}}\nonumber \\
 & \ \ \ +\frac{p'''}{p}-\frac{3p'p''}{p^{2}}+\frac{2p'^{3}}{p^{3}}+\frac{2k}{r^{3}q}-\frac{2}{r^{3}}\\
\lambda''' & =-\frac{q'''}{q}+\frac{3q'q''}{q^{2}}-\frac{kq''}{rq^{2}}-\frac{2q'^{3}}{q^{2}}+\frac{2kq'^{2}}{rq^{3}}+\frac{2kq'}{r^{2}q^{2}}\nonumber \\
 & \ \ \ +\frac{p'''}{p}-\frac{3p'p''}{p^{2}}+\frac{2p'^{3}}{p^{3}}+\frac{2k}{r^{3}q}+\frac{2}{r^{3}}\\
\mu''' & =-\frac{kq''}{rq^{2}}+\frac{2kq'^{2}}{rq^{3}}+\frac{2kq'}{r^{2}q^{2}}+\frac{2k}{r^{3}q}+\frac{4}{r^{3}}
\end{align}
\begin{align}
\nu'''' & =\frac{q''''}{q}-\frac{4q'q'''}{q^{2}}-\frac{kq'''}{rq^{2}}-\frac{3q''^{2}}{q^{2}}+\frac{12q'^{2}q''}{q^{3}}+\frac{6kq'q''}{rq^{3}}\ \ \ \nonumber \\
 & +\frac{3kq''}{r^{2}q^{2}}-\frac{6q'^{4}}{q^{4}}-\frac{6kq'^{3}}{rq^{4}}-\frac{6kq'^{2}}{r^{2}q^{3}}-\frac{6kq'}{r^{3}q^{2}}+\frac{p''''}{p}\nonumber \\
 & -\frac{4p'p'''}{p^{2}}-\frac{3p''^{2}}{p^{2}}+\frac{12p'^{2}p''}{p^{3}}-\frac{6p'^{4}}{p^{4}}-\frac{6k}{r^{4}q}+\frac{6}{r^{4}}\\
\mu'''' & =-\frac{kq'''}{rq^{2}}+\frac{6kq'q''}{rq^{3}}+\frac{3kq''}{r^{2}q^{2}}-\frac{6kq'^{3}}{rq^{4}}-\frac{6kq'^{2}}{r^{2}q^{3}}-\frac{6kq'}{r^{3}q^{2}}\nonumber \\
 & -\frac{6k}{r^{4}q}-\frac{12}{r^{4}}\label{eq:b-2}
\end{align}
\normalsize \vskip8pt

\textbf{Step 3.---}So far we have treated $p(r)$ and $q(r)$ as
two arbitrary functions. We now impose, per \eqref{eq:evol}, that\small
\begin{align}
p'(x) & =\frac{3k^{2}}{4\,r}\frac{p(r)}{q^{2}(r)}\label{eq:b-3}\\
q'(x) & =(1-\Lambda r^{2})\,p(r)\label{eq:ver-1}
\end{align}
\normalsize

From here, we compute the higher derivatives of $p$ and $q$ \emph{recursively
}in terms of $p$ and $q$ themselves, and obtain

\footnotesize

\begin{align}
p'' & =-\frac{3k^{2}}{4r^{2}q^{3}}\Bigl\{2rpq'-rqp'+pq\Bigr\}\\
p''' & =-\frac{3k^{2}}{4r^{3}q^{4}}\Bigl\{2r^{2}pqq''-6r^{2}pq'^{2}\\
 & \ \ \ +4\bigl(r^{2}qp'-rpq\bigr)q'-r^{2}q^{2}p''+2rq^{2}p'-2pq^{2}\Bigr\}
\end{align}
\begin{align}
p'''' & =-\frac{3k^{2}}{4r^{4}q^{5}}\Bigl\{2r^{3}pq^{2}q'''+18\bigl(r^{2}pq-r^{3}qp'\bigr)q'^{2}\\
 & +6\bigl(-3r^{3}pqq'+r^{3}q^{2}p'-r^{2}pq^{2}\bigr)q''+24r^{3}pq'^{3}\\
 & +6\bigl(r^{3}q^{2}p''-2r^{2}q^{2}p'+2rpq^{2}\bigr)q'\\
 & -r^{3}q^{3}p'''+3r^{2}q^{3}p''-6rq^{3}p'+6pq^{3}\Bigr\}
\end{align}
\normalsize and\footnotesize
\begin{align}
q'' & =\bigl(1-\Lambda r^{2}\bigr)p'-2\Lambda rp\\
q''' & =\bigl(1-\Lambda r^{2}\bigr)p''-4\Lambda rp'-2\Lambda p\\
q'''' & =\bigl(1-\Lambda r^{2}\bigr)p'''-6\Lambda rp''-6\Lambda p'\label{eq:b-4}
\end{align}
\normalsize\vskip8pt

\textbf{Step 4.}---Finally, we substitute $\{p',q';p'',q'';...\}$
in \eqref{eq:b-3}--\eqref{eq:b-4} into $\{\nu',\lambda',\mu';\nu'',\lambda'',\mu'';...\}$
in \eqref{eq:b-1}--\eqref{eq:b-2} then into the Christoffel symbols,
the Ricci tensor components, and $\{\mathcal{R},\mathcal{R}',\mathcal{R}''\}$
obtained in \eqref{eq:Chris-00}--\eqref{eq:ddR}. 
\end{document}